\documentclass[]{aa}  

\usepackage{graphicx}
\usepackage{txfonts}
\begin{document}

   \title{\texttt{DeepVel}: deep learning for the estimation of horizontal \\ velocities at the solar surface}


   \author{A. Asensio Ramos\inst{1,2}, I. S. Requerey\inst{1,2}, N. Vitas\inst{1,2}}

   \institute{Instituto de Astrof\'{\i}sica de Canarias, 38205, La Laguna, Tenerife, Spain; \email{aasensio@iac.es}
\and
Departamento de Astrof\'{\i}sica, Universidad de La Laguna, E-38205 La Laguna, Tenerife, Spain
             }

   \date{Received September 15, 1996; accepted March 16, 1997}

 
  \abstract{Many phenomena taking place in the solar photosphere are controlled by plasma motions. 
  Although the line-of-sight component of the velocity can be estimated using the Doppler effect, we do
  not have direct spectroscopic access to the components that are perpendicular to the line-of-sight. These
  components are typically estimated using methods based on local correlation tracking. We have designed
  \texttt{DeepVel}, an end-to-end deep neural network that produces an estimation of the velocity at 
  every single pixel and at every time step and at three different heights in the atmosphere from just two consecutive continuum images. 
  We confront \texttt{DeepVel} with local correlation tracking, pointing out that they give
  very similar results in the time- and spatially-averaged cases. We use the network to study
  the evolution in height of the horizontal velocity field in fragmenting granules, supporting
  the buoyancy-braking mechanism for the formation of integranular lanes in these
  granules. We also show that \texttt{DeepVel} can capture very small vortices, so that
  we can potentially expand the scaling cascade of vortices to very small sizes and durations.}

   \keywords{Sun: granulation, photosphere -- methods: observational, data analysis}
   \authorrunning{Asensio Ramos et al.}
   \titlerunning{\texttt{DeepVel}: deep learning for estimating horizontal velocities}
   \maketitle
%

\section{Introduction}

Motions in the solar photosphere are fundamentally controlled by convection in
a magnetized plasma. The magnetic field topology is controlled by the plasma
motions because the gas pressure is much higher than the magnetic pressure.
As a consequence, many of the phenomena taking place in the photosphere are
dominated by large-, medium- and small-scale plasma motions. Among these phenomena,
an incomplete list would include: emergence of magnetic field thanks to convection, tangling of
magnetic field lines which eventually produces reconnection, convective collapse, cancellation
of magnetic fields, etc. 

Remotely sensing these three-dimensional velocities is important for the
analysis of these events, ideally in combination with spectropolarimetric measurements
to infer the magnetic field. The component along the line of sight (LOS) of the velocity
can be extracted from spectroscopic observations thanks to the Doppler
effect. However, the components of the velocity field in the plane perpendicular
to the LOS cannot be diagnosed spectroscopically.
Different algorithms have been used to trace horizontal flows at the solar surface from  
continuum images \citep{november_simon88,strous95,roudier99,potts04} 
and also magnetograms \citep{kusano02,welsch04,longcope04,schuck05,schuck06,georgoulis06}. 
Among these methods the local correlation tracking \citep[LCT;][]{november_simon88} is the most used one because of its 
simplicity and speed. LCT is a powerful cross-correlation technique for measuring the proper motions of granules. It 
correlates small local windows in several consecutive images to find the best-match displacement. The tracking window is defined 
by a Gaussian function whose full width at half maximum (FWHM) is roughly the size of the features to be tracked. In addition, the spatially localized cross correlation is commonly averaged in time to smooth the transition between consecutive images and reduce the noise induced by atmospheric distortion. All these methods can be considered to give
estimations of the so-called \emph{optical flow}, the vector field that  needs
to be applied to an image to be transformed into a different one.
As such, they might be not strictly representative of the inherent horizontal velocity 
fields.

Given its widespread use, there have been some efforts to
compare the horizontal velocity fields retrieved through LCT with  simulated plasma velocities \citep{rieutord01,matloch10,verma13,yelles14,louis15}. 
Current three-dimensional
magnetohydrodynamical simulations are able to very well reproduce convection in a magnetized plasma, so
one expects that the simulated velocities are a good representation of the real ones in the Sun.
These studies revealed that granules are good tracers for large-scale persistent horizontal flows such as meso- 
and supergranular flows \citep[e.g.,][]{simon88,muller92,derosa04,yelles11,langfellner15} 
or photospheric vortex flows \citep{brandt88,bonet10,vargas11,requerey17}. 
The instantaneous velocity fields—obtained by correlating two consecutive frames—also recover the overall 
morphological features of the flow, but they lack the fine structure observed in the simulated velocities 
\citep{louis15,yelles14}. The correlation increases with the time average \citep{rieutord01,matloch10}
while the LCT-determined horizontal velocities keep being underestimated roughly by a factor of three \citep{verma13}.

In this paper we propose an end-to-end deep learning approach for the estimation of horizontal 
velocity fields in the solar atmosphere based on a deep fully convolutional neural network. The neural 
network is trained on a set of simulated velocity fields. Our approach displays a number of benefits that
clearly overcome existing algorithms by a large margin: it is very fast, uses only two consecutive frames 
and returns the velocity field in every pixel and for every time step. This is done at the expense
of a time consuming training that needs to be done only once.

\section{Deep Neural Networks}
Machine learning is a branch of computer science in which models are directly
extracted from data and not imposed by the researcher. In essence, the majority of machine learning 
techniques can be considered to be nonparametric regression techniques which automatically
adapt to the existing data and also adapt when new data is added.
If these models are sufficiently general, one can apply them to solve complicated
inference problems that cannot be easily solved otherwise. One of the first milestones of 
machine learning was the conception of the perceptron \citep{rosenblatt57}, a very simple
artificial neural network (ANN). Afterwards, ANNs have served
many purposes in machine learning. Specially during the 80's and 90's and thanks to several theoretical developments, ANN
were able to solve problems of increasingly difficulty in supervised and unsupervised regression and
classification. The discovery that ANN with a single hidden
layer are a universal approximant to any nonlinear function \citep{jones90,blum91} allowed them to
be used as a fast substitute on complex inference problems. This was, in large part, facilitated by
the development of the backpropagation algorithm \citep{backpropagation86}, that allowed to train
neural networks using training examples and computing the effect of the difference between the
prediction of the ANN and the training set on the parameters of the network.

ANN had a difficult time during the start of the 21st century 
because of several reasons. First, other techniques with stronger 
theoretical grounds (for instance, support vector machines, Gaussian processes, etc.) allowed
the researchers to understand how the methods were fitting the data and how they can be 
generalized. Second, shallow ANN only allowed to solve relatively simple problems, and once the
networks were made very deep, backpropagation was not able to correctly train them. The reason was
that the gradients with respect to the neural network parameters vanish in deep topologies, so that
training using conjugate gradient stalls. Fortunately, this has radically changed in the last 5 years thanks to some 
breakthroughs. First, it was realized that one of the causes for the failure of backpropagation
in deep architectures was the usage of activation functions (like the usual hyperbolic tangent), that
produced vanishing gradients during backpropagation. This was solved by using activation functions like the Rectified Linear Unit \citep[ReLU;][]{relu10} that we
use in this work, which
do not produce such stalls. Second, fully connected layers were substituted by convolutional layers,
that apply a set of small kernels to the input and give as output the convolution of the input
and the kernels. This induced a reduction in the number of free parameters of the networks without
sacrificing any predictive power. Finally, the appearance of Graphical Processing Units (GPU) on the scene
allowed researchers to train neural networks much faster than it was possible before. This also
opened the possibility to train the networks using huge training sets. This last point can arguably 
be considered the main reason for success of deep learning. Conceptually, deep learning is a set
of machine learning techniques based on learning multiple levels of abstraction of the data. If these
multiple levels are learnt well, deep learning is supposed to generalize well.

\begin{figure}
\includegraphics[width=\columnwidth]{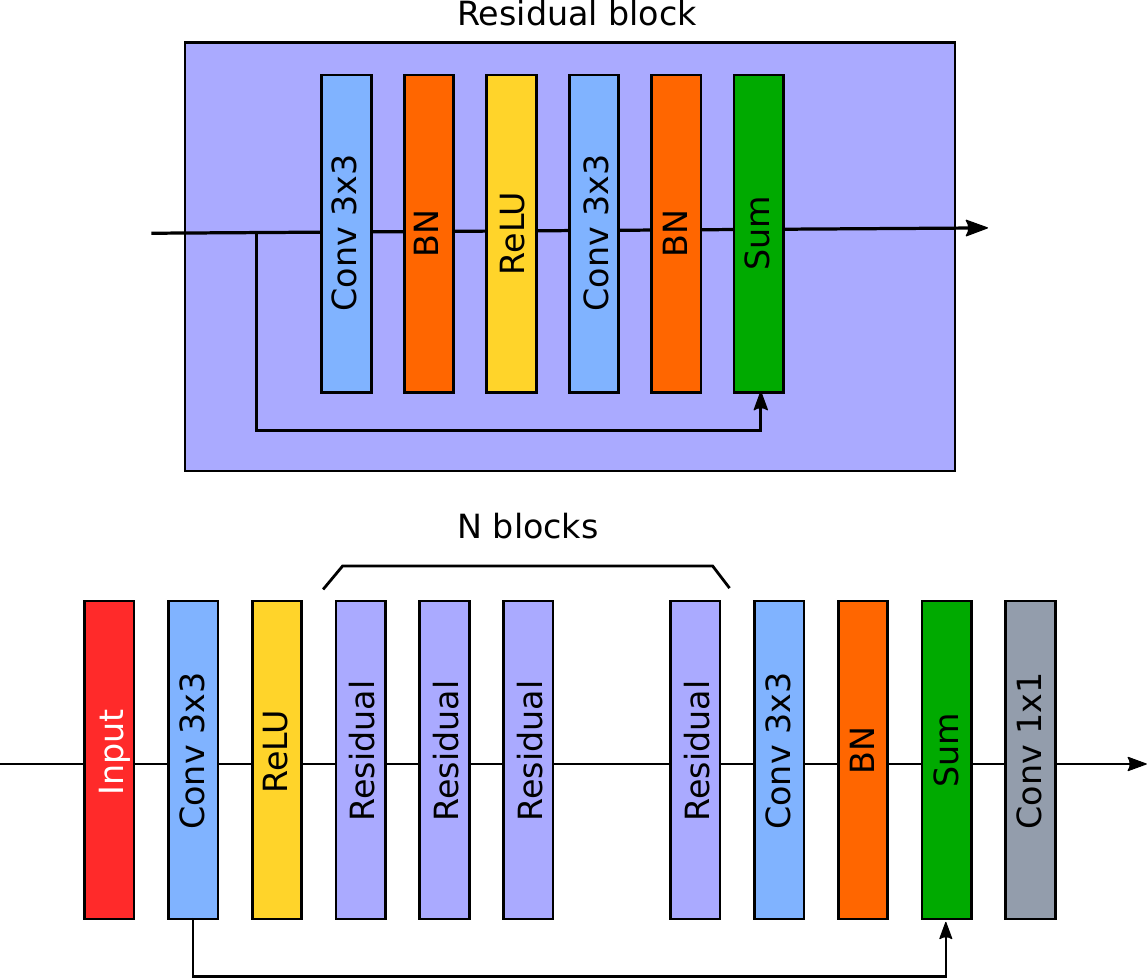}
\caption{Upper panel: residual block. Lower panel: full architecture of the neural network, made of the concatenation
of many residual blocks and a skip connection from the input to the output.  We choose $N=20$ for \texttt{DeepVel}.\label{fig:architecture}}
\end{figure}

In this paper we consider the problem of inferring the horizontal velocity field in the solar
surface from two consecutive continuum images. The case of only two images
represents the worst case scenario and we could have used
more frames for the prediction. However, according to the
results we present in the following, we consider that two frames
gives surprisingly good results.
The end-to-end solution given in this paper, that we term
\texttt{DeepVel}, is a deep neural
network whose topology is described in the following and is trained using velocities extracted
from MHD simulations. 
The network assumes as input two continuum images of size $\mathrm{N}_x \times \mathrm{N}_y$, 
separated by 30 s. The outputs are maps of $v_x$ and $v_y$ at all locations and at three heights in the
atmosphere, corresponding to $\tau_{500}=1, 0.1, 0.01$, with $\tau_{500}$ being the optical depth at 500 nm.
Only the results at $\tau_{500}=1$ can be compared with other algorithms like LCT.

\subsection{Deep Neural Network topology}
The deep network that we use has a fully convolutional architecture, that applies a series 
of convolutions with several small kernels (to be inferred during the training) to the input of every layer. The architecture is graphically
represented in Fig. \ref{fig:architecture}. Each colored rectangle in the figure represents a
different layer, that we describe in the following:
\begin{itemize}
\item \textbf{Input} (red): this layer represents the two input images of size $\mathrm{N}_x \times \mathrm{N}_y$. Consequently, 
this layer represents tensors of size $2 \times \mathrm{N}_x \times \mathrm{N}_y$.
\item \textbf{Conv 3$\times$3} (blue): these layers represent three-dimensional convolutions with a set of 64 
kernels (channels) of size $N_\mathrm{input} \times 3 \times 3$. We keep the number of kernels and their size fixed because they give very good results, with the advantage that
convolutions with $3 \times 3$ kernels can be made very fast in GPUs. The output tensors of these layers
have size $64 \times N_x \times N_y$.
\item \textbf{ReLU} (yellow): these layers represent rectified linear units, which apply the following
operation to every pixel and channel of the input: $\mathrm{ReLU}(x)=x$ if $x \geq 0$ and zero elsewhere.
\item \textbf{BN} (orange): this layer represents batch normalization \citep{batch_normalization15}, a trick used to increase the
convergence speed of the training. It is based on normalizing the input so that it has zero mean
and unit variance, which has been verified to greatly accelerate the training.
\item \textbf{Sum} (green): this layer describes pixelwise addition between the two
inputs.
\item \textbf{Conv 1$\times$1} (grey): this layer defines three-dimensional 
convolution with six $64\times 1 \times 1$ kernels, which is
just a very convenient way to collapse the 64 channels of the last \textbf{Sum} layer of the neural network into
the six velocities that we want to predict. The output tensor of this layer has size $6 \times N_x \times N_y$.
\end{itemize}

As seen from the lower panel of Fig. \ref{fig:architecture}, the network is made of the
concatenation of N so-called \emph{residual} blocks \citep{residual_network16}. We choose
$N=20$ for our implementation and did not carry out any hyperparameter optimization, that we leave
for the future with the aim of optimizing the network. The internal description
of each residual block is displayed in the upper panel of Fig. \ref{fig:architecture}. It
is essentially made of two convolution layers, each one followed by batch normalization layers
and only the first one containing a ReLU activation. Finally, the input of the block is
added pixelwise at the end. The main advantage of the residual blocks is that it accelerates the
training thanks to the skip connection between the input and the output. The output of the
set of residual blocks is then transferred through an additional convolutional layer
with 64 kernels of 3\,$\times$\,3 and a batch normalization layer. The output is then obtained
after convolution with 6 kernels of size $1 \times 1$. The total number of free parameters
of the network is $\sim$1.6\,$\times$\,10$^6$.

\subsection{Training data and training process}
The network is trained using synthetic continuum images from the magneto-convection simulations 
described by \cite{stein12_b} and \cite{stein12_a}.
This simulation box is $\sim$48\,Mm wide in both directions and 20.5\,Mm deep, extending from the temperature minimum 
down to 20\,Mm below the visible surface. The simulated solar
time spans more than an hour in steps of 30 s. The horizontal resolution turns out to be 48\,km, with a total size
of 1008\,$\times$\,1008 pixels. This simulation
displays an appropriate balance between the amount of solar surface simulated and the horizontal resolution.
The \texttt{mhd48-1} snapshots that we use are obtained by advecting a uniform field at the bottom
boundary. This field is increased until it reaches 1\,kG at the bottom boundary and then kept
constant. 

The synthetic images are then treated to simulate a real observation. We choose the Imaging Magnetograph eXperiment \citep[IMaX;][]{imax11} on board the \textsc{Sunrise} balloon borne observatory \citep{sunrise10,sunrise11} as a target. \textsc{Sunrise} has a telescope of 1 m diameter and the images that IMaX
provides have a spatial sampling of 39.9\,km. It is interesting to note that
the spatial sampling of the simulated images used in the training and those of IMaX do not
exactly coincide (48\,km vs. 39.9\,km). However, we demonstrate later than 
an appropriately trained network will generalize correctly independently of the size of the structures.

Given that \textsc{Sunrise} was a balloon mission that observed at a height
of $\sim$ 40 km above the Earth surface, the observations are barely affected by the atmosphere. One of the reasons
to choose IMaX in our tests is that the instrument
has provided long time series of very high-quality diffraction-limited images in both
flights \citep[e.g.,][]{lagg_imax10,marian11,sunrise17}, which have been used often for LCT studies 
\citep[e.g.,][]{bonet10,yelles11,requerey14,requerey17}. 
We simulate the effect of IMaX following 
the approach of \cite{asensio_phasediv_imax12}, which is based on the
detailed analysis of \cite{santiago_vargas09}. We consider telescope aberrations up
to 45 Zernike modes. The amplitudes are considered to be normally distributed with
diagonal covariance and a total rms wavefront error (WFE) amounting to $\lambda/9$. 
These telescope aberrations can be considered to be constant during an observation, so we keep them fixed.
The remaining atmosphere is accounted for by considering a wavefront with turbulent Kolmogorov statistics \citep{noll76} with a
rms WFE of $\lambda/9$. Although these perturbations are very specific for \textsc{Sunrise}/IMaX, we think that
the neural network trained with these images can be safely applied (or perhaps easily re-trained for
different instrumental configurations).

From the available three-dimensional volume of 
1008\,$\times$\,1008 continuum images for all
timesteps, we randomly extract two patches of 
50$\times$50 pixels in the same spatial position and separated by 30 s in time. A total of 30000 such
pairs are randomly selected as input for the training. The outputs are the six 50\,$\times$\,50 images containing
$v_x$ and $v_y$ at $\tau_{500}=1,0.1$ and 0.01, respectively, for each timestep. We also generate another 
set of 1000 samples using the same strategy, which is used as a validation set 
to avoid overfitting. These are used during training to check
that the deep network generalizes correctly and is not memorizing 
the training set.

The neural network has been developed using the \texttt{Keras} Python library, with the 
\texttt{Tensorflow} backend for the computations. All the training is compiled by \texttt{Tensorflow}
to be run in the NVIDIA Tesla K40 and Titan X GPUs\footnote{The trained neural network ready to be applied
to solar images, together with the infrastructure to train it with different simulations
can be found in \texttt{https://github.com/aasensio/deepvel}.}.
The training is carried out by minimizing the squared difference between the output of the network
and the velocities in the training set. It is known that optimizing the $\ell_2$ norm
of the difference might lead to too smooth predictions, which is not appropriate
for typical uses of deep networks for machine vision in natural images. In the last
few years, improvements in this direction have been performed using a second deep network
that is used to measure the quality of the prediction. Both networks are trained
as a generative adversarial network \citep[GAN;][]{goodfellow14}, which results in impressive
results \citep{ledig16}. We know from 
the simulations that the horizontal velocity fields are relatively
smooth, so we stick with the simpler $\ell_2$ norm for our case. We leave
the analysis of using GANs for the training for the future.

All inputs are normalized to the median intensity of the quiet Sun (which also needs 
to be done once the network is applied to real observations)
and velocities
are normalized to the interval $[0,1]$ using the minimum and maximum velocities in the training set.
The optimization is done with the Adam stochastic first-order gradient-based optimization algorithm 
\citep{adam14} with a learning rate $\epsilon=10^{-4}$. As in any stochastic optimization method,
the gradient is estimated from subsets of the input samples, 
also known as
batches. We use batches of 32 samples and train the network for 30 epochs, where an epoch is finished
once all training samples have been used. Therefore, the number of iterations is then 900000.

\begin{figure*}
\includegraphics[width=\textwidth]{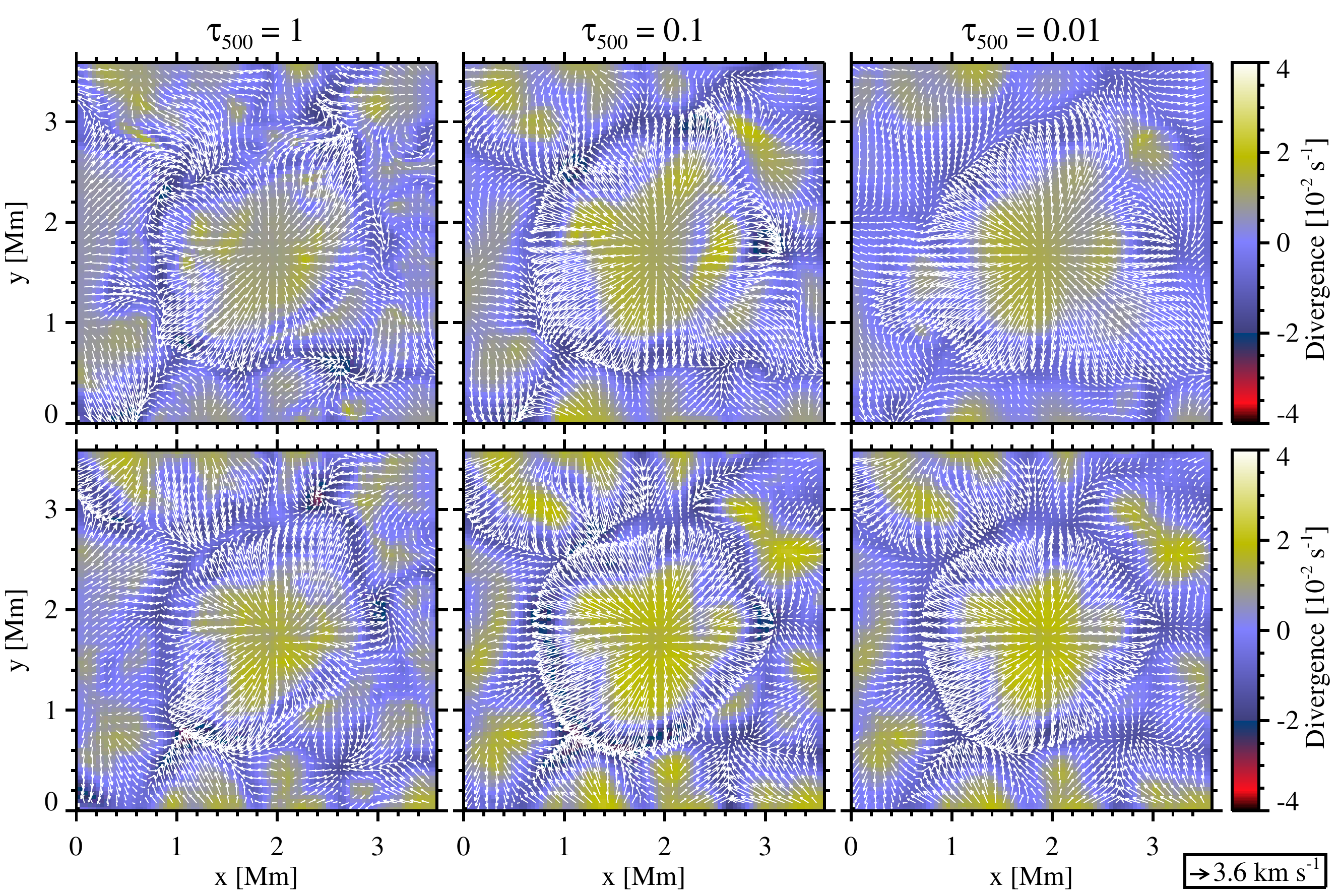}
\caption{Instantaneous horizontal velocity field (white arrows) and divergence maps (background images) at three heights in the atmosphere, corresponding to $\tau_{500}=1, 0.1, 0.01$, 
for \texttt{MANCHA} (upper row) and \texttt{DeepVel} (bottom row) velocities.}
\label{fig:SimDeepVel}
\end{figure*}

\subsection{Validation}
In absence of a technique similar to \texttt{DeepVel} that can be
applied to observations, we validate the method using 3D magneto-convection
simulations carried out with the \texttt{MANCHA} code \citep{felipe10,khomenko17}. 
The extent of the simulation domain is 24 Mm $\times$ 24 Mm in the horizontal plane 
and 1.4 Mm vertically, with 1152 grid cells in each horizontal direction and 102 uniformly spaced grid points in the vertical direction.
The domain is open for the mass flows at the bottom boundary and closed at the top. 
The radiative transfer losses are computed assuming local thermodynamical equilibrium with precomputed opacities.
The magnetic field is initiated through the Biermann battery and amplified by 
the local dynamo \citep[similarly to][]{vogler07}. 
The snapshots used in this study were taken when the total magnetic 
field reached 10 G at the unit optical depth.
The synthetic continuum maps are degraded so that the pixel size is
equivalent to those of the training set. We use two consecutive
continuum maps to infer the horizontal velocity field and compare it 
with the ones extracted from the simulations. The upper panels of
Fig. \ref{fig:SimDeepVel} display the velocity fields of the simulations
at three different optical heights in the atmosphere. The underlying map
is the divergence of the horizontal field, which 
is computed as $\nabla\,\mathbf{v}=\partial v_x/
\partial x+\partial v_y / \partial y$ for $\mathbf{v}=(v_x,v_y)$. Note that  
positive values indicate diverging 
flows, while negative values point to converging flows.
The lower panels display the results of \texttt{DeepVel}, which very
nicely reproduce the results from the simulation, specially for
$\tau_{500}=1, 0.1$. The results for $\tau_{500}=0.01$ are
slightly less similar although the general appearance
is still valuable. Figure \ref{fig:SimDeepVel} only shows a small field of view (FOV), but similar results are found for the rest of the simulated field. Specifically, 
the velocity field vectors of the whole simulated FOV have a Pearson linear correlation coefficient of 
0.82, 0.85, and 0.76 for $\tau_{500}=1,$ 0.1, and 0.01, respectively. Additionally, $v_x$ displays correlation
coefficients of 0.82, 0.84, and 0.75 for the same values of $\tau_{500}$, while the figures turn out to
be 0.83, 0.86, and 0.78 for $v_y$ for the same $\tau_{500}$ heights. We consider that this experiment validates \texttt{DeepVel}.

\begin{figure*}
\includegraphics[width=\textwidth]{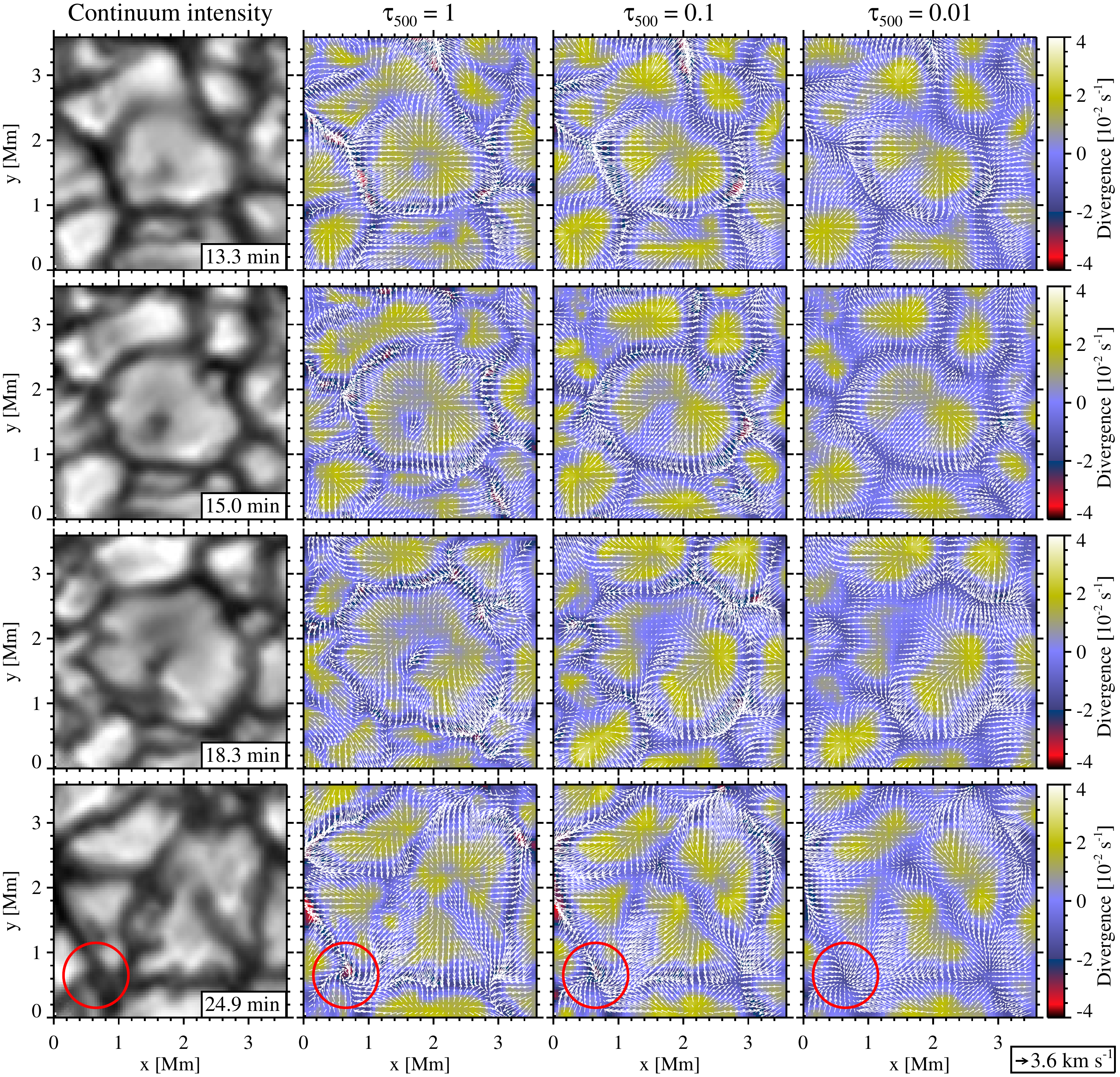}
\caption{Evolution of granules as seen in close-ups of continuum intensity maps (left column) and the instantaneous horizontal velocity field (white arrows) and divergence maps (background images) at three heights in the atmosphere (right columns), corresponding to $\tau_{500}=1, 0.1, 0.01$. The area is located within the black solid rectangle in Fig. \ref{fig:fov}(a). Red circles (with a radius of 0.5 Mm) in the bottom row indicate the location of a small-scale vortex flow. Elapsed time is given in the bottom-right corner of each continuum image. The whole evolution is presented in a movie provided as online material. The arrow at the lower right corner can be used to visually estimate
the amplitude of the velocity field. The whole evolution is presented in a movie provided as online material.}
\label{fig:granule}
\end{figure*}

\section{Results}
Once the network is trained, we apply it to real IMaX observations 
from the first \textsc{Sunrise} flight. The observational data were obtained 
on 2009 June 9 from 01:30:54 to 02:02:29 UT, in a quiet-Sun region close to disk center. 
The 31.6\,min length dataset has a temporal cadence of 33.25\,s.
We point out that the temporal cadence is 
larger than the one used in the training, so that the network might
slightly overestimate the velocities\footnote{This could have been 
alleviated if the simulated snapshots would have been obtained 
with the IMaX temporal cadence. We did not follow this path
because we wanted to work with public simulations that are available
for anyone willing to re-train \texttt{DeepVel}.}. We also note that the spatial resolution
of the observations is also slightly better than those of the 
simulations, but we do not expect large effects. Even though the training was 
done with images of size 50\,$\times$\,50, given the fully convolutional character
of the network, we apply it seamlessly to the full FOV of the
instrument, that amounts to 736\,$\times$\,736 pixels (29.3\,$\times$\,29.3\,Mm$^{2}$). 
The computing time is $\sim$2\,s per image using a Titan X GPU, and an order
of magnitude larger if the computation is done in a CPU.

\subsection{Inferred velocity fields}
In Fig. \ref{fig:granule} and the movie in the online 
material\footnote{The movie can also be obtained from the code 
repository}, we display the inferred horizontal velocity field for a small portion of the FOV at four different time steps. 
The first column shows the continuum image, while the
rest of columns display the divergence of the horizontal velocity 
field at the three different heights in the atmosphere, together with
the instantaneous vector field. Note that the results
displayed in Fig. \ref{fig:granule} are impossible to be obtained
using LCT due to the somehow large spatial and temporal smearing windows
that need to be used to increase the correlation and produce robust results.

In general, velocities in lower layers tend to be larger in absolute value and also 
with larger spatial complexity. Although horizontal diverging flows are 
similar at the three heights considered, stronger converging flows are 
seen in intergranular lanes in deeper layers. Additionally, the horizontal
size of these zones of converging flows is much smaller in deep layers.

\begin{figure*}
\includegraphics[width=\textwidth]{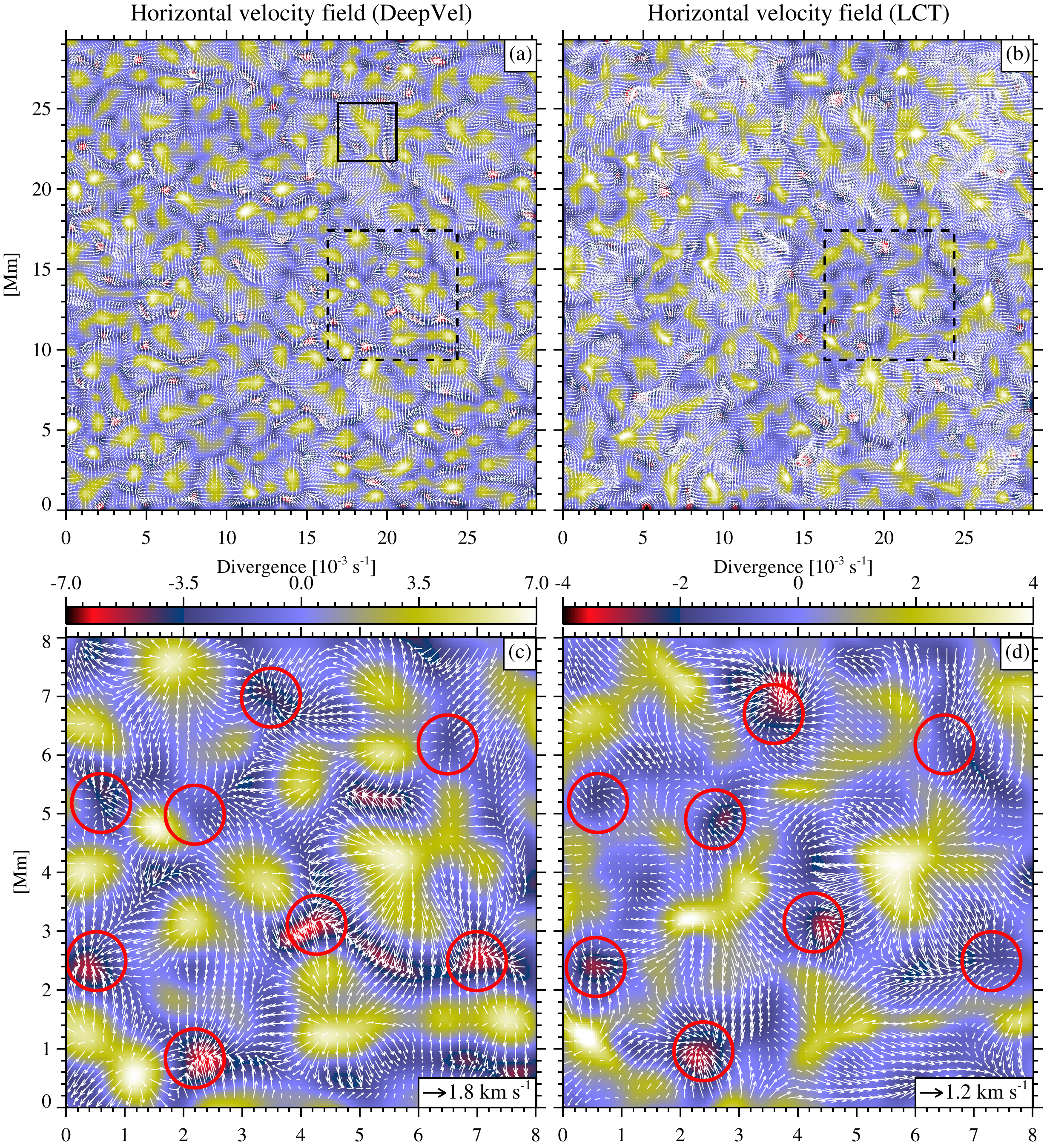}
\caption{Average horizontal velocity field (white arrows) and divergence maps (background images) for the FOV of IMaX computed by \texttt{DeepVel} (a) and LCT (b), respectively. The black solid box marks the region analyzed in Fig. \ref{fig:granule}. Close-ups of the black dashed rectangles are shown in panels (c) and (d) for \texttt{DeepVel} and LCT, respectively. Red circles represent the locations of converging flows.}
\label{fig:fov}
\end{figure*}

\begin{figure*}
\includegraphics[width=\textwidth]{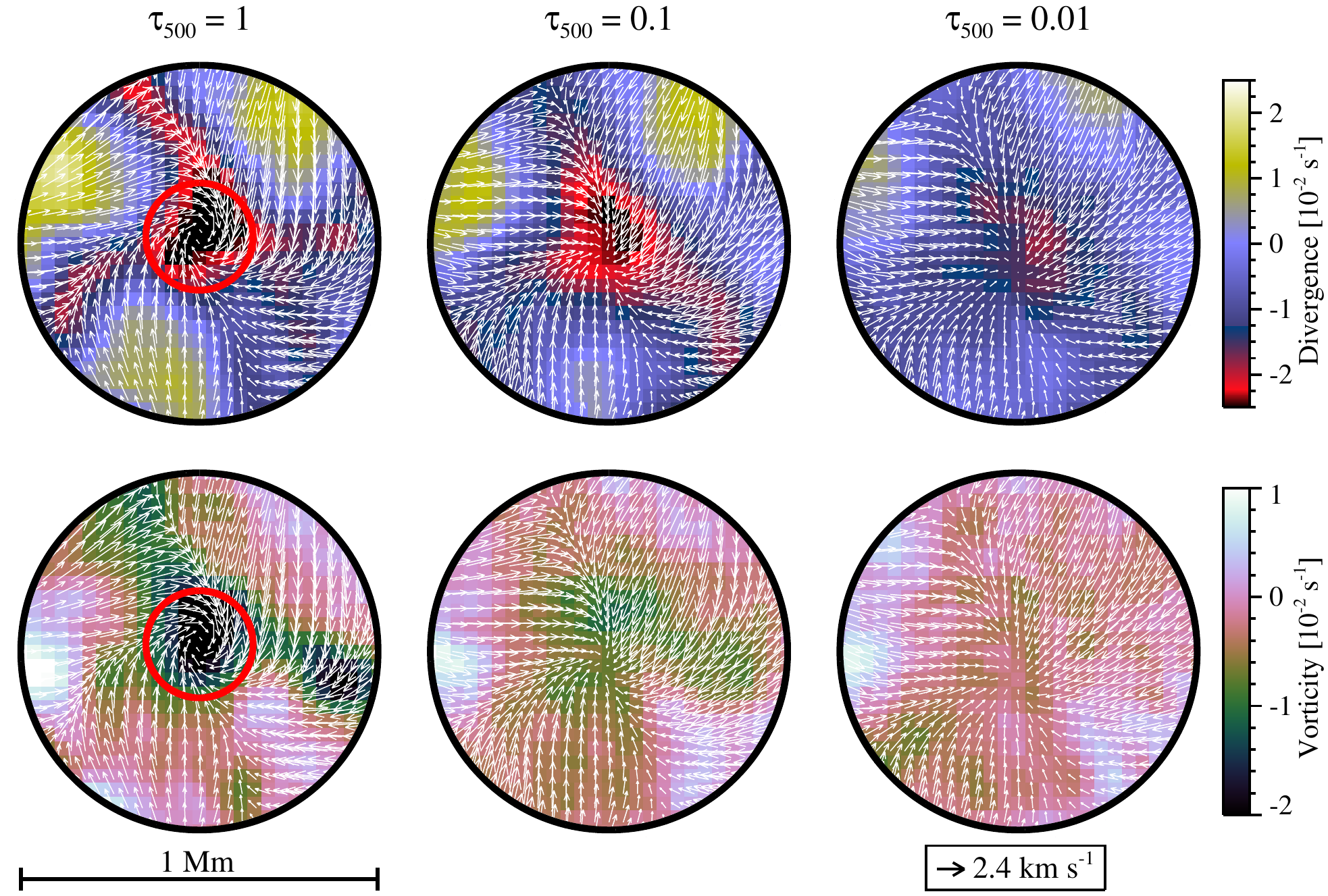}
\caption{Close-up of the small-scale vortex flow shown in Fig. \ref{fig:granule} by red circles. The upper panels
show the divergence of the horizontal velocity field at the three selected heights, while the
lower panels display the vertical vorticity. Red circles (with a radius of 150 km) show the size of the small-scale vortex flow.}
\label{fig:vortex}
\end{figure*}

We find very interesting the behavior observed in some granules at $t=15$\,min, 
like the one at positions $(x,y)=(0.8,2.6)$\,Mm. This granule appear to be formed by 
the aggregation
of two or more smaller portions. The continuum image of the
granule shows a slightly dark lane separating bright regions.
This structure is also clearly seen in the velocity field at $\tau_{500}=0.1$
and $\tau_{500}=0.01$, while it disappears for deeper layers. It looks
like these dark lanes in the middle of granules are a
consequence of a converging flow in upper layers that does
not reach very deep. Interestingly, the velocity fields before ($t=13.3$ min)
and after ($t=18.3$\,min) help us understand what is happening. We are 
witnessing a fragmenting granule that is being divided into two by a converging
velocity field taking place at higher layers, which later propagates
towards lower layers. Note that at $t=13.3$ min, the division is barely
visible in the intensity image, but it is already present at $\tau_{500}=0.01$.
Finally, at $t=18.3$\,min, the granule is divided in two parts, with a clear
intergranule between them. Our observations seem to be compatible with the
buoyancy-braking mechanism \citep{massaguer_zahn80,ploner99}. In this model, 
the gas above large granules reduce their upward velocity because of the 
increase in the mass in upper layers. It looses buoyancy and catastrophically
collapses forming a dark lane if energy losses cannot be compensated.
According to our observations, it also develops strong converging velocity
flows when eventually forming a new intergranular lane.
The fact that \texttt{DeepVel} shows this behavior means that this mechanism 
has to be the predominant one in the simulations and that the behavior in
upper layers is connected to what is going on at lower layers.

Other instances of fragmenting granules and appearance of dark structures
inside bright granules can be seen in the observations. They all appear
to share a similar behavior, except for the specific details. For instance, 
the dark spot at $(x,y)=(1.6,1.4)$ Mm at $t=15$ min seems to reach lower
layers slightly faster than the previous example.

\subsection{Comparison with LCT}

In order to test the performance of \texttt{DeepVel}, we compare it with the well known 
LCT algorithm. The LCT technique recovers horizontal proper motions by tracking 
intensity features in continuum images. We use a Gaussian tracking window with 
an FWHM\,=\,600\,km and then average the cross-correlation function over 
30\,min. We confront this velocity field with that obtained by \texttt{DeepVel} 
at $\tau_{500}=1$, temporally averaged over 30\,min. In addition, the vector field is 
spatially averaged with the LCT tracking window size. The top panels of 
Fig. \ref{fig:fov} show the horizontal velocity arrows and the corresponding 
divergence maps obtained by \texttt{DeepVel} (Fig. \ref{fig:fov}a)  and LCT (Fig. \ref{fig:fov}b) for 
the whole FOV of IMaX. We find that the \texttt{DeepVel} velocities 
are $1.15$ times larger in magnitude than the LCT ones, which almost perfectly
coincides with the overestimation expected because measurements in IMaX are taken
every 33.25 s, instead of the 30 s used in the training (the correction factor would
be ($\times$1.11). The velocity fields have Pearson linear correlation coefficients of 
0.81 and 0.84 for $v_x$ and $v_y$, respectively, while it goes down to 0.8 for the whole velocity
vector. Additionally, the correlation coefficient for the divergence is $0.71$.

This correlation is already evident from the visual inspection of the flow and divergence maps. 
The bottom panels of Fig. \ref{fig:fov} display an enlarged view of a smaller 
region, marked by a black dashed rectangle in the upper panels. Even though there 
exist some morphological differences, both maps show the same flow features. In particular, 
they display an equivalent mesogranular pattern, which is revealed through positive 
divergence structures on scales between granulation and supergranulation. Such a cellular 
pattern is commonly found in both observations and hydrodynamical simulations when the 
LCT technique is applied to intensity images
\citep{simon88,muller92,roudier98,matloch10,yelles11,requerey17}. At the junction of
mesogranular cells, smaller structures such as converging flows
\citep{bonet10,vargas11,requerey17} are also observed. The locations of the convergence centers
are marked by red circles in Fig. \ref{fig:fov}(c) and (d). Despite their small size, they are
equally retrieved in both velocity fields.

The time-averaging increases the correlation between the simulated plasma velocities and the 
LCT ones \citep{rieutord01,matloch10}. The same applies for the correlation between the 
smoothed \texttt{DeepVel} velocities and the LCT flows. Specifically, we get correlation coefficients 
of 0.71, 0.75, 0.79, and 0.80 for averaging times of 5, 10, 20, and 30\,min, respectively. 
This results from the fact that the \texttt{DeepVel} velocities are a reliable representation 
of the instantaneous horizontal flow fields, while the LCT velocities are only comparable to 
plasma velocities typically at time scales longer than the average granule lifetime.

\subsection{Small-scale vortex flow}

Small-scale vortex flows have been first detected as swirling motions of bright points \citep{bonet08}, and later through LCT \citep{bonet10,vargas11,requerey17}. They have a diameter of $\sim$\,1\,Mm \citep{bonet08,bonet10,vargas11}, their lifetime varies from 5 to 20\,min \citep{bonet08,bonet10}, and they appear located at mesogranular junctions \citep{requerey17}. The spatial distribution and size of such vortices is shown by red circles in Fig. \ref{fig:fov}(d). The temporal and spatial scales are larger than those expected from simulations, where the vortices have lifetimes of only a few minutes ($\sim$\,3.5\,min) and diameters of  $\sim$\,100\,km \citep{moll11}. This differences are likely due to the smoothing produced by the time and spatial average of the LCT technique.

Figure \ref{fig:vortex} shows the velocity fields for a close-up of the vortex flow marked in
Fig. \ref{fig:granule} by red circles. The underlying maps in the upper panels show the divergence, while the lower panels display the vertical vorticity of the horizontal velocity field, defined as 
$(\nabla \times \mathbf{v})_z=\partial v_y/
\partial x-\partial v_x / \partial y$. The vortex has a very small size, surely
smaller than 300\,km in diameter (see red circle in Fig. \ref{fig:vortex}), and lasts for a very short time, in the range 30-60 s, because
it is only clearly visible in one time step, and can be guessed in the previous and next frames. 

The vortex flow has a central zone with a 
very strong negative divergence at $\tau_{500}=1$, reaching values up to $-0.03$ s$^{-1}$, which are
more than an order of magnitude larger than the median value found
by \cite{requerey17} as a consequence of the much smaller size. The same behavior
is seen in the vorticity, with values that also reach $-0.03$ s$^{-1}$ at $\tau_{500}=1$, negative
meaning clockwise rotation. This value of vorticity is an order of magnitude larger than that detected with LCT \citep{bonet10,vargas11,requerey17}, but comparable to that found in simulations \citep[e.g.,][]{Kitiashvili11}.

Interestingly, the \texttt{DeepVel} results show a different picture of
the vortex flow at higher layers. First, material is still strongly advected towards
the center of the vortex, with divergences that are still of the order of $-0.02$ s$^{-1}$, while
the vorticity becomes much smaller in higher layers. This behavior resembles that of the “bathtub effect” \citep{Nordlund85}, in which the circular velocity is amplified as the plasma contracts with depth.

\section{Conclusions}
We have developed \texttt{DeepVel}, an end-to-end approach for the estimation of 
instantaneous and per-pixel
horizontal velocity fields based on a deep network. The network is fully
convolutional, so that it can be applied to input images of arbitrary size, providing
outputs of exactly the same size. In addition, it is very fast, has no parameters
to be tuned and is available to the community as open source. Concerning speed, we
note that it can be improved with some architectural changes, but we think that it 
is already fast enough for our standards.

We have checked that the 
spatially and temporally averaged horizontal velocity field 
provided by the network is very similar to that obtained with LCT. However, the power of 
\texttt{DeepVel} is that this same information can also be obtained instantaneously, contrary
to LCT. Additionally, we provide the velocity field at three different heights in the
atmosphere, something that might look counterintuitive at a first look. It is clear from
the results presented here (both from simulations and observations) that the network is able to
correctly generalize and is not overfitting.

Similar to this application, we expect deep learning to be increasingly applied to
Solar Physics as more high-quality data is obtained and needs to be analyzed. An incomplete
list of possible potential applications in which we are already working on would include
fast image deconvolution, fast 2D and 3D spectropolarimetric inversions, fast control
of adaptive optics, etc.

\begin{acknowledgements}
We thank B. Ruiz Cobo \& F. J. de Cos Juez for very useful
comments on an early version of the paper. We also thank R. Abreu for
initial discussions on the subject of deep learning.
Financial support by the Spanish Ministry of Economy and Competitiveness 
through projects AYA2014-60476-P Consolider-Ingenio 2010 CSD2009-00038
and ESP2014-56169-C6 are gratefully acknowledged. AAR also acknowledges financial support
through the Ram\'on y Cajal fellowships. We also thank the NVIDIA Corporation for the donation
of the Titan X GPU used in this research.
We acknowledge PRACE for awarding us access to resource MareNostrum based in Barcelona/Spain.
This research has made use of NASA's Astrophysics Data System Bibliographic Services.
We acknowledge the community effort devoted to the development of the following open-source packages that were
used in this work: \texttt{numpy} (\texttt{numpy.org}), \texttt{matplotlib} (\texttt{matplotlib.org}), \texttt{Keras} (\texttt{https://keras.io}), and \texttt{Tensorflow} {\texttt{http://www.tensorflow.org}}.
\end{acknowledgements}


\begin{thebibliography}{56}
\expandafter\ifx\csname natexlab\endcsname\relax\def\natexlab#1{#1}\fi

\bibitem[{{Asensio Ramos} {et~al.}(2012){Asensio Ramos}, {Mart{\'{\i}}nez
  Gonz{\'a}lez}, {Khomenko}, \& {Mart{\'{\i}}nez
  Pillet}}]{asensio_phasediv_imax12}
{Asensio Ramos}, A., {Mart{\'{\i}}nez Gonz{\'a}lez}, M.~J., {Khomenko}, E., \&
  {Mart{\'{\i}}nez Pillet}, V. 2012, \aap, 539, A42

\bibitem[{{Barthol} {et~al.}(2011){Barthol}, {Gandorfer}, {Solanki},
  {Sch{\"u}ssler}, {Chares}, {Curdt}, {Deutsch}, {Feller}, {Germerott},
  {Grauf}, {Heerlein}, {Hirzberger}, {Kolleck}, {Meller}, {M{\"u}ller},
  {Riethm{\"u}ller}, {Tomasch}, {Kn{\"o}lker}, {Lites}, {Card}, {Elmore},
  {Fox}, {Lecinski}, {Nelson}, {Summers}, {Watt}, {Mart{\'{\i}}nez Pillet},
  {Bonet}, {Schmidt}, {Berkefeld}, {Title}, {Domingo}, {Gasent Blesa}, {Del
  Toro Iniesta}, {L{\'o}pez Jim{\'e}nez}, {{\'A}lvarez-Herrero},
  {Sabau-Graziati}, {Widani}, {Haberler}, {H{\"a}rtel}, {Kampf}, {Levin},
  {P{\'e}rez Grande}, {Sanz-Andr{\'e}s}, \& {Schmidt}}]{sunrise11}
{Barthol}, P., {Gandorfer}, A., {Solanki}, S.~K., {et~al.} 2011, \solphys, 268,
  1

\bibitem[{{Blum} \& {Li}(1991)}]{blum91}
{Blum}, E.~K. \& {Li}, L.~K. 1991, Neural Networks, 4, 511

\bibitem[{{Bonet} {et~al.}(2008){Bonet}, {M{\'a}rquez}, {S{\'a}nchez Almeida},
  {Cabello}, \& {Domingo}}]{bonet08}
{Bonet}, J.~A., {M{\'a}rquez}, I., {S{\'a}nchez Almeida}, J., {Cabello}, I., \&
  {Domingo}, V. 2008, ApJL, 687, L131

\bibitem[{{Bonet} {et~al.}(2010){Bonet}, {M{\'a}rquez}, {S{\'a}nchez Almeida},
  {Palacios}, {Mart{\'{\i}}nez Pillet}, {Solanki}, {del Toro Iniesta},
  {Domingo}, {Berkefeld}, {Schmidt}, {Gandorfer}, {Barthol}, \&
  {Kn{\"o}lker}}]{bonet10}
{Bonet}, J.~A., {M{\'a}rquez}, I., {S{\'a}nchez Almeida}, J., {et~al.} 2010,
  \apjl, 723, L139

\bibitem[{{Brandt} {et~al.}(1988){Brandt}, {Scharmer}, {Ferguson}, {Shine}, \&
  {Tarbell}}]{brandt88}
{Brandt}, P.~N., {Scharmer}, G.~B., {Ferguson}, S., {Shine}, R.~A., \&
  {Tarbell}, T.~D. 1988, \nat, 335, 238

\bibitem[{{De Rosa} \& {Toomre}(2004)}]{derosa04}
{De Rosa}, M.~L. \& {Toomre}, J. 2004, \apj, 616, 1242

\bibitem[{{Felipe} {et~al.}(2010){Felipe}, {Khomenko}, \&
  {Collados}}]{felipe10}
{Felipe}, T., {Khomenko}, E., \& {Collados}, M. 2010, \apj, 719, 357

\bibitem[{{Georgoulis} \& {LaBonte}(2006)}]{georgoulis06}
{Georgoulis}, M.~K. \& {LaBonte}, B.~J. 2006, \apj, 636, 475

\bibitem[{{Goodfellow} {et~al.}(2014){Goodfellow}, {Pouget-Abadie}, {Mirza},
  {Xu}, {Warde-Farley}, {Ozair}, {Courville}, \& {Bengio}}]{goodfellow14}
{Goodfellow}, I.~J., {Pouget-Abadie}, J., {Mirza}, M., {et~al.} 2014, ArXiv
  e-prints [\eprint[arXiv]{1406.2661}]

\bibitem[{He {et~al.}(2016)He, Zhang, Ren, \& Sun}]{residual_network16}
He, K., Zhang, X., Ren, S., \& Sun, J. 2016, in 2016 {IEEE} Conference on
  Computer Vision and Pattern Recognition, {CVPR} 2016, Las Vegas, NV, USA,
  June 27-30, 2016, 770--778

\bibitem[{Ioffe \& Szegedy(2015)}]{batch_normalization15}
Ioffe, S. \& Szegedy, C. 2015, in Proceedings of the 32nd International
  Conference on Machine Learning (ICML-15), ed. D.~Blei \& F.~Bach (JMLR
  Workshop and Conference Proceedings), 448--456

\bibitem[{{Jones}(1990)}]{jones90}
{Jones}, L.~K. 1990, in {Proceedings of the IEEE}, {78}, 1585

\bibitem[{{Khomenko} {et~al.}(2017){Khomenko}, {Vitas}, {Collados}, \& {de
  Vicente}}]{khomenko17}
{Khomenko}, E.~V., {Vitas}, N., {Collados}, M., \& {de Vicente}, A. 2017,
  submitted to A\&A

\bibitem[{Kingma \& Ba(2014)}]{adam14}
Kingma, D.~P. \& Ba, J. 2014, CoRR, abs/1412.6980

\bibitem[{{Kitiashvili} {et~al.}(2011){Kitiashvili}, {Kosovichev}, {Mansour},
  \& {Wray}}]{Kitiashvili11}
{Kitiashvili}, I.~N., {Kosovichev}, A.~G., {Mansour}, N.~N., \& {Wray}, A.~A.
  2011, \apjl, 727, L50

\bibitem[{{Kusano} {et~al.}(2002){Kusano}, {Maeshiro}, {Yokoyama}, \&
  {Sakurai}}]{kusano02}
{Kusano}, K., {Maeshiro}, T., {Yokoyama}, T., \& {Sakurai}, T. 2002, \apj, 577,
  501

\bibitem[{{Lagg} {et~al.}(2010){Lagg}, {Solanki}, {Riethm{\"u}ller},
  {Mart{\'i}nez Pillet}, {Sch{\"u}ssler}, {Hirzberger}, {Feller}, {Borrero},
  {Schmidt}, {del Toro Iniesta}, {Bonet}, {Barthol}, {Berkefeld}, {Domingo},
  {Gandorfer}, {Kn{\"o}lker}, \& {Title}}]{lagg_imax10}
{Lagg}, A., {Solanki}, S.~K., {Riethm{\"u}ller}, T.~L., {et~al.} 2010, ApJl,
  723, L164

\bibitem[{{Langfellner} {et~al.}(2015){Langfellner}, {Gizon}, \&
  {Birch}}]{langfellner15}
{Langfellner}, J., {Gizon}, L., \& {Birch}, A.~C. 2015, \aap, 581, A67

\bibitem[{Ledig {et~al.}(2016)Ledig, Theis, Huszar, Caballero, Aitken, Tejani,
  Totz, Wang, \& Shi}]{ledig16}
Ledig, C., Theis, L., Huszar, F., {et~al.} 2016, CoRR, abs/1609.04802

\bibitem[{{Longcope}(2004)}]{longcope04}
{Longcope}, D.~W. 2004, \apj, 612, 1181

\bibitem[{{Louis} {et~al.}(2015){Louis}, {Ravindra}, {Georgoulis}, \&
  {K{\"u}ker}}]{louis15}
{Louis}, R.~E., {Ravindra}, B., {Georgoulis}, M.~K., \& {K{\"u}ker}, M. 2015,
  \solphys, 290, 1135

\bibitem[{{Mart{\'i}nez Gonz{\'a}lez} {et~al.}(2011){Mart{\'i}nez
  Gonz{\'a}lez}, {Asensio Ramos}, {Manso Sainz}, {Khomenko}, {Mart{\'i}nez
  Pillet}, {Solanki}, {L{\'o}pez Ariste}, {Schmidt}, {Barthol}, \&
  {Gandorfer}}]{marian11}
{Mart{\'i}nez Gonz{\'a}lez}, M.~J., {Asensio Ramos}, A., {Manso Sainz}, R.,
  {et~al.} 2011, ApJL, 730, L37+

\bibitem[{{Mart{\'i}nez Pillet} {et~al.}(2011){Mart{\'i}nez Pillet}, {Del Toro
  Iniesta}, {{\'A}lvarez-Herrero}, {Domingo}, {Bonet}, {Gonz{\'a}lez
  Fern{\'a}ndez}, {L{\'o}pez Jim{\'e}nez}, {Pastor}, {Gasent Blesa}, {Mellado},
  {Piqueras}, {Aparicio}, {Balaguer}, {Ballesteros}, {Belenguer}, {Bellot
  Rubio}, {Berkefeld}, {Collados}, {Deutsch}, {Feller}, {Girela}, {Grauf},
  {Heredero}, {Herranz}, {Jer{\'o}nimo}, {Laguna}, {Meller}, {Men{\'e}ndez},
  {Morales}, {Orozco Su{\'a}rez}, {Ramos}, {Reina}, {Ramos}, {Rodr{\'i}guez},
  {S{\'a}nchez}, {Uribe-Patarroyo}, {Barthol}, {Gandorfer}, {Knoelker},
  {Schmidt}, {Solanki}, \& {Vargas Dom{\'i}nguez}}]{imax11}
{Mart{\'i}nez Pillet}, V., {Del Toro Iniesta}, J.~C., {{\'A}lvarez-Herrero},
  A., {et~al.} 2011, {Sol. Phys.}, 268, 57

\bibitem[{{Massaguer} \& {Zahn}(1980)}]{massaguer_zahn80}
{Massaguer}, J.~M. \& {Zahn}, J.-P. 1980, \aap, 87, 315

\bibitem[{{Matloch} {et~al.}(2010){Matloch}, {Cameron}, {Shelyag}, {Schmitt},
  \& {Sch{\"u}ssler}}]{matloch10}
{Matloch}, {\L}., {Cameron}, R., {Shelyag}, S., {Schmitt}, D., \&
  {Sch{\"u}ssler}, M. 2010, \aap, 519, A52

\bibitem[{{Moll} {et~al.}(2011){Moll}, {Cameron}, \& {Sch{\"u}ssler}}]{moll11}
{Moll}, R., {Cameron}, R.~H., \& {Sch{\"u}ssler}, M. 2011, \aap, 533, A126

\bibitem[{{Muller} {et~al.}(1992){Muller}, {Auffret}, {Roudier}, {Vigneau},
  {Simon}, {Frank}, {Shine}, \& {Title}}]{muller92}
{Muller}, R., {Auffret}, H., {Roudier}, T., {et~al.} 1992, \nat, 356, 322

\bibitem[{Nair \& Hinton(2010)}]{relu10}
Nair, V. \& Hinton, G.~E. 2010, in Proceedings of the 27th International
  Conference on Machine Learning (ICML-10), June 21-24, 2010, Haifa, Israel,
  807--814

\bibitem[{{Noll}(1976)}]{noll76}
{Noll}, R.~J. 1976, Journal of the Optical Society of America, 66, 207

\bibitem[{{Nordlund}(1985)}]{Nordlund85}
{Nordlund}, A. 1985, \solphys, 100, 209

\bibitem[{{November} \& {Simon}(1988)}]{november_simon88}
{November}, L.~J. \& {Simon}, G.~W. 1988, \apj, 333, 427

\bibitem[{{Ploner} {et~al.}(1999){Ploner}, {Solanki}, \& {Gadun}}]{ploner99}
{Ploner}, S.~R.~O., {Solanki}, S.~K., \& {Gadun}, A.~S. 1999, \aap, 352, 679

\bibitem[{{Potts} {et~al.}(2004){Potts}, {Barrett}, \& {Diver}}]{potts04}
{Potts}, H.~E., {Barrett}, R.~K., \& {Diver}, D.~A. 2004, \aap, 424, 253

\bibitem[{{Requerey} {et~al.}(2014){Requerey}, {Del Toro Iniesta}, {Bellot
  Rubio}, {Bonet}, {Mart{\'{\i}}nez Pillet}, {Solanki}, \&
  {Schmidt}}]{requerey14}
{Requerey}, I.~S., {Del Toro Iniesta}, J.~C., {Bellot Rubio}, L.~R., {et~al.}
  2014, \apj, 789, 6

\bibitem[{{Requerey} {et~al.}(2017){Requerey}, {Del Toro Iniesta}, {Bellot
  Rubio}, {Mart{\'{\i}}nez Pillet}, {Solanki}, \& {Schmidt}}]{requerey17}
{Requerey}, I.~S., {Del Toro Iniesta}, J.~C., {Bellot Rubio}, L.~R., {et~al.}
  2017, \apjs, 229, 14

\bibitem[{{Rieutord} {et~al.}(2001){Rieutord}, {Roudier}, {Ludwig}, {Nordlund},
  \& {Stein}}]{rieutord01}
{Rieutord}, M., {Roudier}, T., {Ludwig}, H.-G., {Nordlund}, {\AA}., \& {Stein},
  R. 2001, \aap, 377, L14

\bibitem[{{Rosenblatt}(1957)}]{rosenblatt57}
{Rosenblatt}, F. 1957, Cornell Aeronautical Laboratory Report, 85, 460

\bibitem[{{Roudier} {et~al.}(1998){Roudier}, {Malherbe}, {Vigneau}, \&
  {Pfeiffer}}]{roudier98}
{Roudier}, T., {Malherbe}, J.~M., {Vigneau}, J., \& {Pfeiffer}, B. 1998, \aap,
  330, 1136

\bibitem[{{Roudier} {et~al.}(1999){Roudier}, {Rieutord}, {Malherbe}, \&
  {Vigneau}}]{roudier99}
{Roudier}, T., {Rieutord}, M., {Malherbe}, J.~M., \& {Vigneau}, J. 1999, \aap,
  349, 301

\bibitem[{{Rumelhart} {et~al.}(1986){Rumelhart}, {Hinton}, \&
  {Williams}}]{backpropagation86}
{Rumelhart}, D.~E., {Hinton}, G.~E., \& {Williams}, R.~J. 1986, Nature, 323,
  533

\bibitem[{{Schuck}(2005)}]{schuck05}
{Schuck}, P.~W. 2005, \apjl, 632, L53

\bibitem[{{Schuck}(2006)}]{schuck06}
{Schuck}, P.~W. 2006, \apj, 646, 1358

\bibitem[{{Simon} {et~al.}(1988){Simon}, {Title}, {Topka}, {Tarbell}, {Shine},
  {Ferguson}, {Zirin}, \& {SOUP Team}}]{simon88}
{Simon}, G.~W., {Title}, A.~M., {Topka}, K.~P., {et~al.} 1988, \apj, 327, 964

\bibitem[{{Solanki} {et~al.}(2010){Solanki}, {Barthol}, {Danilovic}, {Feller},
  {Gandorfer}, {Hirzberger}, {Riethm{\"u}ller}, {Sch{\"u}ssler}, {Bonet},
  {Mart{\'i}nez Pillet}, {del Toro Iniesta}, {Domingo}, {Palacios},
  {Kn{\"o}lker}, {Bello Gonz{\'a}lez}, {Berkefeld}, {Franz}, {Schmidt}, \&
  {Title}}]{sunrise10}
{Solanki}, S.~K., {Barthol}, P., {Danilovic}, S., {et~al.} 2010, ApJL, 723,
  L127

\bibitem[{{Solanki} {et~al.}(2017){Solanki}, {Riethm{\"u}ller}, {Barthol},
  {Danilovic}, {Deutsch}, {Doerr}, {Feller}, {Gandorfer}, {Germerott}, {Gizon},
  {Grauf}, {Heerlein}, {Hirzberger}, {Kolleck}, {Lagg}, {Meller}, {Tomasch},
  {van Noort}, {Blanco Rodr{\'{\i}}guez}, {Gasent Blesa}, {Balaguer
  Jim{\'e}nez}, {Del Toro Iniesta}, {L{\'o}pez Jim{\'e}nez}, {Orozco
  Su{\'a}rez}, {Berkefeld}, {Halbgewachs}, {Schmidt}, {{\'A}lvarez-Herrero},
  {Sabau-Graziati}, {P{\'e}rez Grande}, {Mart{\'{\i}}nez Pillet}, {Card},
  {Centeno}, {Kn{\"o}lker}, \& {Lecinski}}]{sunrise17}
{Solanki}, S.~K., {Riethm{\"u}ller}, T.~L., {Barthol}, P., {et~al.} 2017, ArXiv
  e-prints [\eprint[arXiv]{1701.01555}]

\bibitem[{{Stein}(2012)}]{stein12_a}
{Stein}, R.~F. 2012, Living Reviews in Solar Physics, 9, 4

\bibitem[{{Stein} \& {Nordlund}(2012)}]{stein12_b}
{Stein}, R.~F. \& {Nordlund}, {\AA}. 2012, \apjl, 753, L13

\bibitem[{{Strous}(1995)}]{strous95}
{Strous}, L.~H. 1995, in ESA Special Publication, Vol. 376, Helioseismology,
  213

\bibitem[{{Vargas Dom{\'i}nguez}(2009)}]{santiago_vargas09}
{Vargas Dom{\'i}nguez}, S. 2009, PhD thesis, Universidad de La Laguna, La
  Laguna

\bibitem[{{Vargas Dom{\'{\i}}nguez} {et~al.}(2011){Vargas Dom{\'{\i}}nguez},
  {Palacios}, {Balmaceda}, {Cabello}, \& {Domingo}}]{vargas11}
{Vargas Dom{\'{\i}}nguez}, S., {Palacios}, J., {Balmaceda}, L., {Cabello}, I.,
  \& {Domingo}, V. 2011, \mnras, 416, 148

\bibitem[{{Verma} {et~al.}(2013){Verma}, {Steffen}, \& {Denker}}]{verma13}
{Verma}, M., {Steffen}, M., \& {Denker}, C. 2013, \aap, 555, A136

\bibitem[{{V{\"o}gler} \& {Sch{\"u}ssler}(2007)}]{vogler07}
{V{\"o}gler}, A. \& {Sch{\"u}ssler}, M. 2007, \aap, 465, L43

\bibitem[{{Welsch} {et~al.}(2004){Welsch}, {Fisher}, {Abbett}, \&
  {Regnier}}]{welsch04}
{Welsch}, B.~T., {Fisher}, G.~H., {Abbett}, W.~P., \& {Regnier}, S. 2004, \apj,
  610, 1148

\bibitem[{{Yelles Chaouche} {et~al.}(2014){Yelles Chaouche}, {Moreno-Insertis},
  \& {Bonet}}]{yelles14}
{Yelles Chaouche}, L., {Moreno-Insertis}, F., \& {Bonet}, J.~A. 2014, \aap,
  563, A93

\bibitem[{{Yelles Chaouche} {et~al.}(2011){Yelles Chaouche}, {Moreno-Insertis},
  {Mart{\'{\i}}nez Pillet}, {Wiegelmann}, {Bonet}, {Kn{\"o}lker}, {Bellot
  Rubio}, {del Toro Iniesta}, {Barthol}, {Gandorfer}, {Schmidt}, \&
  {Solanki}}]{yelles11}
{Yelles Chaouche}, L., {Moreno-Insertis}, F., {Mart{\'{\i}}nez Pillet}, V.,
  {et~al.} 2011, \apjl, 727, L30

\end{thebibliography}

\end{document}